\documentclass{jfm}
\usepackage{graphicx}
\usepackage{epstopdf, epsfig}
\usepackage{xcolor}
\usepackage{MnSymbol}
\usepackage{wasysym}

\shorttitle{Self similar flapping foil wakes}
\shortauthor{N. S. Lagopoulos, G. D. Weymouth, B. Ganapathisubramani}

\title{Universal scaling law for drag-to-thrust \\ wake transition in flapping foils}

\author{N. S. Lagopoulos\aff{1}
  \corresp{\email{N.Lagopoulos@soton.ac.uk}},
  G. D. Weymouth\aff{2}
 \and B. Ganapathisubramani\aff{1}}

\affiliation{\aff{1}Aerodynamics and Flight Mechanics Group,
University of Southampton, UK
\aff{2}Southampton Marine and Maritime Institute,
University of Southampton, UK}

\begin{document}

\maketitle

\begin{abstract}
Reversed $\emph{von}~K\acute{a}rm\acute{a}n$ streets are responsible for a velocity surplus in the wake of flapping foils, indicating the onset of thrust generation. However, the wake pattern cannot be predicted based solely on the flapping peak-to-peak amplitude $A$ and frequency $f$ because the transition also depends sensitively on other details of the kinematics. In this work we replace $A$ with the cycle-averaged swept trajectory $\mathcal{T}$ of the foil chord-line. Two dimensional simulations are performed for pure heave, pure pitch and a variety of heave-to-pitch coupling. In a phase space of dimensionless $\mathcal{T}-f$ we show that the drag-to-thrust wake transition of all tested modes occurs for a modified Strouhal $St_{\mathcal{T}}\sim 1$. Physically the product $\mathcal{T}\cdot f$ expresses the induced velocity of the foil and indicates that propulsive jets occur when this velocity exceeds $U_{\infty}$. The new metric offers a unique insight into the thrust producing strategies of biological swimmers and flyers alike as it directly connects the wake development to the chosen kinematics enabling a self similar characterisation of flapping foil propulsion.
\end{abstract}


\section{Introduction}  \label{intro}
Almost all aquatic and flying animals generate thrust via the oscillatory motion of foil-like body parts e.g. tail, fin etc. Moreover flapping foil systems are often associated with high efficiency and strong side forces, ideal for manoeuvring \citep{Read2003}. Thus, many studies have focused on the analysis and implementation of these biological configurations into man made designs \citep{Fish2006,Triantafyllou2004, Wang2005} although the underlying physics is still not
clearly understood.

\begin{figure}
  \centerline{\includegraphics[width =5.5 in]{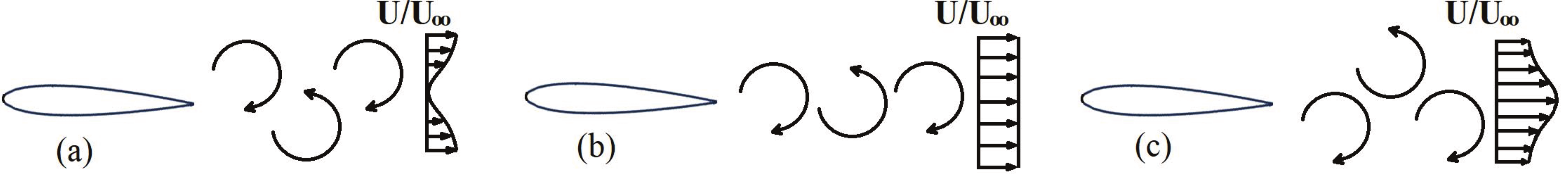}}
   \caption{Drag-to-Thrust transition within the wake of a symmetric flapping foil. (a) BvK street, (b) Neutral line, (c) reversed BvK wake.}\label{figure1}
\end{figure}

Here we aim to determine the drag-to-thrust wake transition of these flapping mechanisms via the analysis of the vortex pattern development. Assuming foil undulations above the $\emph{lock in}$ frequency \citep{Thiria2006, Vial2004} we observe at least three basic wake patterns \citep{vonKarman1935}: the classic $B\acute{e}nard~\emph{von}~K\acute{a}rm\acute{a}n$ (Bvk) street where $U_{wake} < U_{\infty}$ (figure \ref{figure1}a), the neutral line where $U_{wake} \sim U_{\infty}$ (figure \ref{figure1}b) and the reversed BvK wake, where $U_{wake} > U_{\infty}$ (figure \ref{figure1}c). The latter is synonymous to the drag-to-thrust wake transition although a lag exists between this phenomenon and the foil's overall transition towards thrust. This is due to the fact that a weak velocity surplus cannot overcome profile drag or velocity fluctuations and pressure differences within the control volume\citep{Streitlien1998, Ramamurti2001, Bohl2009}. 

As the driving factors of BvK reversal we typically consider the oscillating amplitude and the oscillating frequency $f$ of the kinematics \citep{Koochesfahani1989} . The former is expressed by the trailing-edge (TE) peak-to-peak amplitude $A$. In dimensionless terms it is often normalised by the thickness D or the chord length $\mathcal{C}$ of the foil. In a similar fashion the frequency is often expressed as a reduced frequency $k = U_{\infty}/(f \mathcal{C})$ \citep{Birnbaum1924}, a thickness based Strouhal number $Sr = (fD)/U_{\infty}$ \citep{Diana2008a} or a chord length based Strouhal number $St_{\mathcal{C}} = 1/k$ \citep{Cleaver2012}. \cite{Triantafyllou1991} suggested a modified amplitude based Strouhal number $St_A = (fA)/U_{\infty}$. By including both the frequency and the amplitude of oscillation, $St_A$ can potentially characterise the BvK reversal by a single factor as opposed to $k$, $Sr$ and $St_{\mathcal{C}}$. Studies of \cite{Anderson1998} and \cite{Read2003} showed that optimal efficiency occurs for a short range of $St_A \sim [0.2, 0.4]$. This was also supported by \cite{Taylor2003} and \cite{Triantafyllou1993} who observed that the majority of natural fliers and swimmers prefer to cruise within this range. According to \cite{Andersen2017}, BvK reversal occurs at different $St_A$ values for pure heave and pure pitch. Therefore, $St_A$ cannot be regarded as an expression of self similarity.
 
The fundamental problem is that characterizing the motion only by the tail amplitude fails to capture the contribution of the other points of the foil. This becomes important when the heaving component is significant, resulting in the generation of strong $\emph{leading edge vortices}$ (LEV) which travel downstream and blend with the $\emph{trailing edge vortices}$ (TEV). Instead, we need to take into account the length of the entire path travelled in a period rather than the maximum distance from equilibrium expressed by $A_D$.

In this study, we formulate a novel length scale, which characterizes the Bvk reversal of harmonically flapping foils. Two-dimensional simulations are conducted at a Reynolds number of $Re =1173$ for a rigid NACA0016 profile and three basic harmonic kinematics: pure heave, pure pitch and heave-pitch coupling. Additional higher Reynolds number simulations ($Re=11730$) are used to quantify the Reynolds number effects. The influence of different pivot points is examined for pure pitch and coupled motions. In addition, we analyse the impact of different maximum effective angles of attack on coupled kinematics. Finally, we develop a new metric based on the chord-wise averaged path travelled by the foil, in order to determine BvK wake reversal for a vast range of harmonic motions.

\subsection{Geometry and Kinematics}
We consider a rigid NACA0016 with a thickness $D = 0.16\mathcal{C}$. The foil performs simple harmonic oscillations around a stationary equilibrium position, against a uniform free stream velocity $U_x = U_{\infty}$. The lateral direction of every point along this chordline is denoted with $y(t, s)$. Here $t$ is the time and $s$ is the $\mathcal{C}$-normalised coordinate along the chord ranging from 0 at LE to 1 at TE. Pure pitch is modelled as a sinusoidal rotation about a specified pivot point along the chordline ($s = \mathcal{P}$, where $\mathcal{P}$ is the non-dimensional distance between the LE and the pivot point along the chordline) and pure heave as a sinusoidal lateral translation of the entire chordline. Coupled motion occurs by the superposition of these pure motions: 

\begin{equation}
\begin{array}{l}
\displaystyle ~~~~~~~~~~~~~~~~~~~~~y_c(t) = y_ h(t) + y_ {\theta}(t)~~~where:\\
\\
\displaystyle  y_h(t) =  h_\mathit{0} \sin(2f\pi t)~~~~,~~~~y_ {\theta}(t) =(1-\mathcal{P})\mathcal{C}\sin({\theta}(t))~~~and:\\
\\
\displaystyle~~~~~~~~~~~~~~~~~~~~~\theta(t) = \theta_\mathit{0} \sin(2f\pi t +\psi)\\ 
\end{array} 
\label{eq:xdef}
\end{equation}

Here $\theta(t)$ is the instantaneous value of pure pitch whilst $h_{0}$ and $\theta_{0}$ are the amplitudes of pure heave and pure pitch respectively. The phase difference between pitch and heave is expressed as $\psi$. Here $\psi$ = $90^{\circ}$ because this value is considered optimal in terms of propulsive efficiency \citep{Platzer2008}.

Another important kinematic parameter in coupled motions is the $\emph{effective}$ angle of attack $\alpha_\emph{eff}(t)$ which is the summation of the instantaneous pitch angle $\theta(t)$ and the heave induced angle of attack. Thus for $\psi$ = $90^{\circ}$ the amplitude of $\alpha_\emph{eff}(t)$ is:

\begin{equation}
\alpha = \arctan \frac{\dot{h_{0}}}{U_\infty} - \theta_{0}
\end{equation}
where $\dot{h_{0}}$ is the amplitude of $\dot{y_h(t)}$ = $dy_h/dt$. In this study we differentiate coupled motions by varying $\mathcal{P}$ and $\alpha$ while keeping $\psi$ = $90^{\circ}$. 

\begin{figure}
  \centerline{\includegraphics[width =4 in]{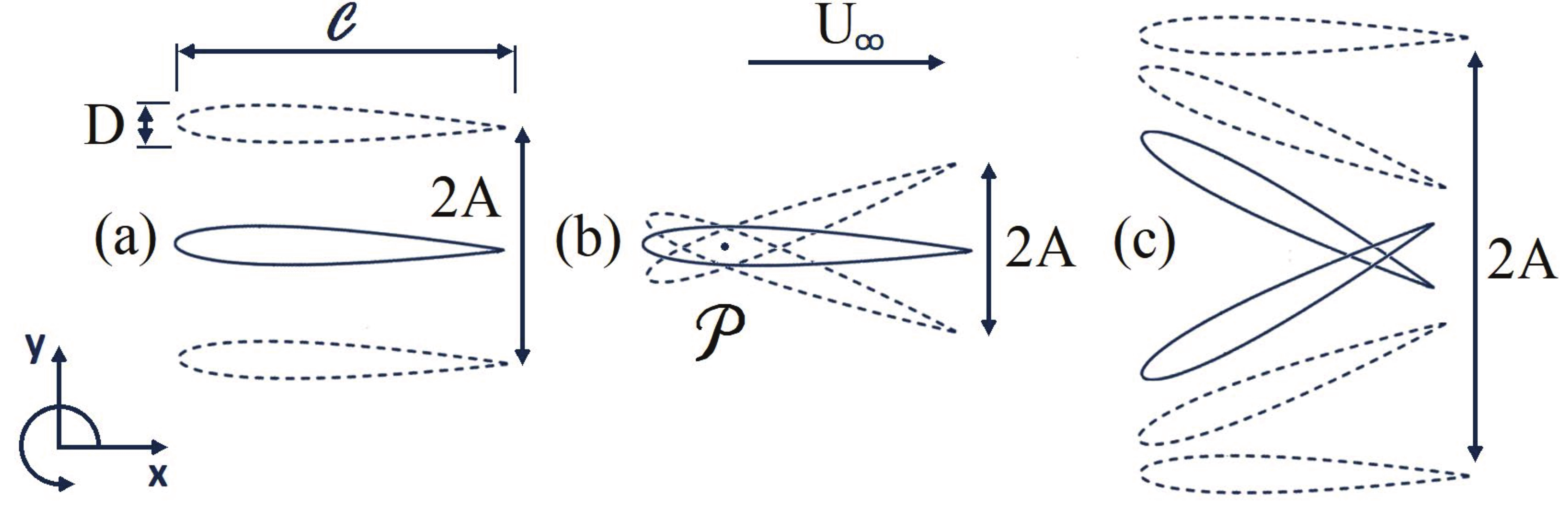}}
   \caption{Foil kinematics, geometry and coordinate system. (a) pure heave , (b) pure pitch , (c) coupled motion.}\label{figure2}
\end{figure}

\subsection{Dimensionless parameters}
Three non-dimensional parameters are used to describe the interaction between the
oscillating foil and the free stream: the Reynolds number based on chord length, $Re= U_{\infty} \mathcal{C} / \nu$ (where $\nu$ is the kinematic viscosity) the thickness based Strouhal number, $Sr$ and the non-dimensional TE amplitude ($A_D$):

\begin{equation}
Sr = \frac{D f}{U_{\infty}}~~,~~A_D =\frac{2 y_{t}(t_{max})}{D}
\end{equation}

Thus $St_{A} = Sr \cdot A_D$ and can be understood as the ratio between the speed of the foil tip and $U_{\infty}$ \citep{Diana2008a}. 

The thrust coefficient $C_t$ is expressed by simply normalising the total force acting on
the $x-\emph{axis}$ by the dynamic pressure of the freestream. Time averaged quantities are presented with an overbar to distinguish them from their instantaneous counterparts. 

\begin{equation}
C_t= \frac{F_x}{\frac{1}{2}\rho U_\infty ^2 c} 
\end{equation}

\subsection{Computational Method}\label{sec:filetypes}
The CFD solver chosen for this study can simulate complex geometries and moving
boundaries for a wide range of Reynolds numbers in 2D and 3D domains, by utilizing
the boundary data immersion method BDIM, \citep{Weymouth2011}. BDIM solves
the viscous time-dependent Navier-Stokes equations and simulates the entire domain
by combining the moving body and the ambient fluid through a kernel function. This
technique has quadratic convergence and has been validated for flapping foil simulations over a wide range of kinematics \citep{Maertens2015,Polet2015}.

The mesh configuration is a rectangular Cartesian grid with a dense uniform grid near the body and in the near wake, and exponential grid stretching used in the far-field and the numerical domain uses a uniform inflow, zero-gradient outflow and free-slip conditions on the upper and lower boundaries.Furthermore no slip boundary conditions are used on the oscillating foil. Mesh density is expressed in terms of grid points per chord. A uniform grid of $\delta x = \delta y = \mathcal{C}/192$ is used for the results in this work based on the results of the convergence study shown in table \ref{tab:gr1}.

\begin{table}
\begin{center}
\def~{\hphantom{0}}
 \begin{tabular}{c c c c} 
 Grid density & Thrust coefficient & $|\Delta C_t|$ & Relative $|\Delta C_t|$ \% \\ [3pt] 
  064 & 0.494 & 9.3E -2  & 15.8\\
  128 & 0.555 & 3.2E -2  & 5.45\\ 
  192 & 0.575 & 1.2E -2  & 2.04\\
  256 & 0.580 & 0.7E -2  & 1.19\\
  512 & 0.587 & ---  & --- \\ 
  \end{tabular}
 \caption{Computational statistics of grid convergence for a harmonically flapping foil.}
\label{tab:gr1}
\end{center}
\end{table}

\begin{figure}
  \centerline{\includegraphics[width =4.3 in]{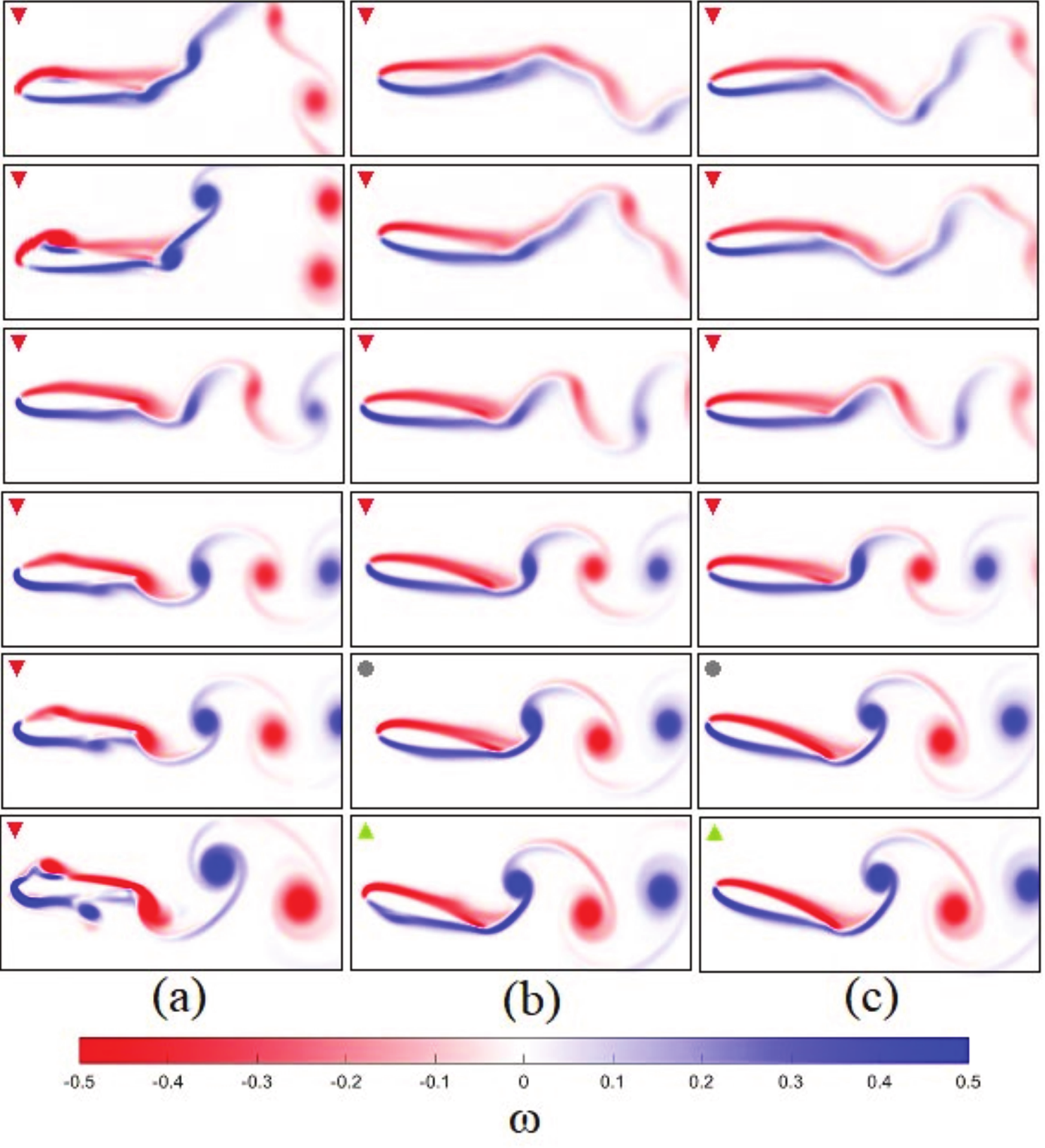}}
  \caption{Contour plots of normalised instantaneous vorticity magnitude of the 30th cycle at $Sr~=~0.12$ and $Re~=~1173$, for (a) pure heave at $A_D \sim [0.5-0.85]$, (b) pitch at $A_D \sim[0.82-1.17]$ and (c) coupled motion at $A_D \sim[1.02-1.37]$. Drag regime $\color{red}{\filledmedtriangledown}$, neutral state $\color{gray}{\CIRCLE}$, thrust producing flow $\color{green}{\filledmedtriangleup}$.}\label{figure3}
\end{figure}

\section{Results and Discussion}

\begin{figure}
  \centerline{\includegraphics[width =5 in]{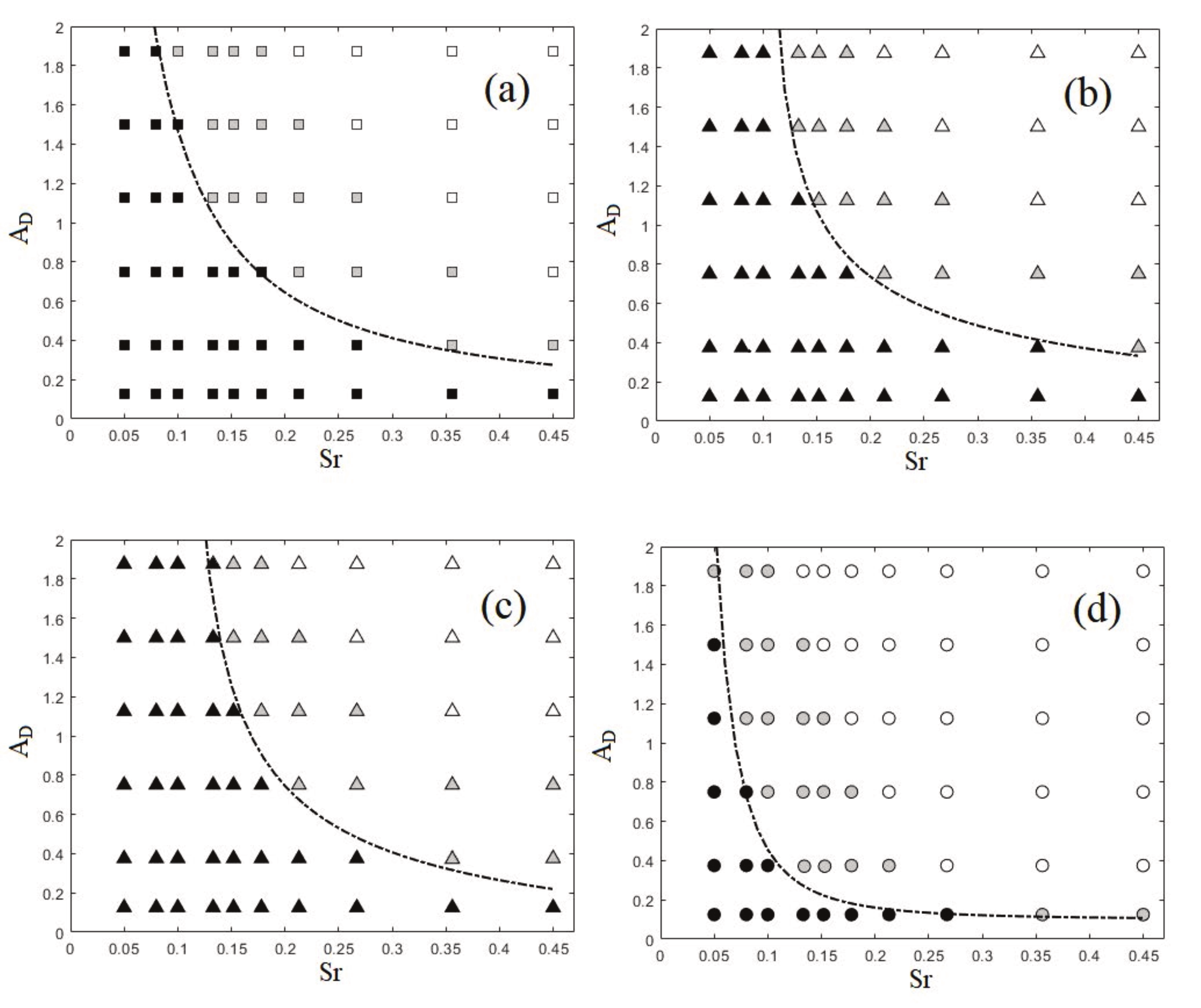}}
  \caption{$A_D-Sr$ wake map for various kinematics at $Re~=~1173$ for (a) pure heave, (b) pure pitch at $\mathcal{P}=0.25$, (c) pure pitch at $\mathcal{P}=0$ and (\textit{d}) coupled motion at $\mathcal{P}=0.25$ and $\alpha=10^{\circ}$ .Black dots: BvK street. Grey dots: reversed BvK wake. White
dots: wake symmetry breaking. The dashed black curve corresponds to the neutral line.}
\label{figure4}

\end{figure}

\subsection{Wake comparison of different kinematics}
 According to \cite{Diana2008b}, the Reynolds number range of naturally occurring flapping foils is $100<Re<10000$. In this study, most simulations are conducted for $Re=1173$ to be within this range and to enable comparison with the experiments of \cite{Diana2008a}. Additional simulations are conducted for selected cases at $Re=11730$ to examine the higher Reynolds number effects. The pivot points tested for pure pitch and coupling are $\mathcal{P} = 0$ and $0.25$. The coupled motion is also tested for three values of $\alpha = 5^{\circ},10^{\circ}$  and $ 20^{\circ}$.
 
We analyse the wake patterns and resultant hydrodynamic loads for the above mentioned kinematics. Various stages of the wake development can be seen in figure 3 for pure heave, pure pitch for $\mathcal{P} = 0.25$ and coupled motion for $\mathcal{P} = 0.25$ and $\alpha = 10^{\circ}$. The transition from
the BvK (third row) to the neutral wake where vortices are shed in-line (fourth row) and later the reversed BvK (fifth and sixth rows) is in agreement with literature \citep{Koochesfahani1989,Diana2008a,Andersen2017} . At lower $Sr-A_D$ combinations more complicated wake patterns e.g. 2P wakes \citep{Williamson1988} at the first row of figure 3 in accordance with those observed by \cite{Andersen2017} for wedge type foils. At such low $Sr-A_D$ combinations coupled motions are dominated by one of the two modes e.g. for $A_D = 0.4 , Sr = 0.1$, the heaving contribution to the foil displacement is less than $5\%$ for all the coupled cases studied. Therefore, a coupled motion around this region is acting more like a pure pitching case. 

Among the three kinematic test cases significant discrepancies can be seen in close proximity to the foil, most notably at the LE. The deep stall (high $\delta \alpha / \delta t$ across the chord) experienced by the pure heaving foil generates LEVs of sizes comparable to the TEVs which travel across the chord and blend with the wake. A closer look at figure \ref{figure3}c reveals that the coupled motion generates the smallest amount of dynamic separation among the three cases. Finally as seen in the last row of figure  \ref{figure3} even when the BvK street is fully reversed, some the cases exist within the drag producing regime. This lag is expected since a weak propulsive wake is not enough to overcome the profile drag or to compensate for the velocity fluctuations and pressure differences that exist within the control volume \citep{Streitlien1998,Ramamurti2001, Bohl2009}.

Figure \ref{figure4} shows the best fit curve that isolates the neutral line (where $U_{wake} \sim U_{\infty}$), for the different harmonics. These best fit curves are reproduced in figure \ref{figure5}a in order to compare this neutral line across different kinematics. Although $Sr-A_D$ phase diagram is suitable to examine a specific kinematics, it is clear that this does not universally describe wake transitions. This is the result of the unique interactions between LEVs and TEVs that are specific to the motion type. Consequently this demonstrates the need for a self similar classification of the oscillating amplitude to accurately determine wake development.

\subsection{An Alternative Length Scale}

Fundamentally, the foil generates thrust force by displacing and accelerating fluid out of its path as it moves through its prescribed trajectory. The quantity of fluid displaced is dependent on the product of the chord length $\mathcal{C}$ times the path length travelled by the foil over one period of oscillation. Thus,a proper indicator of  the wake's drag-to-thrust transition should reflect the length of the curve traversed by the foil within a cycle. 

To quantify the aforementioned distance, we compare different length approximations of the path length ($\mathcal{L}$) covered by the TE in one non-dimensional period for a heaving foil. We estimate this length in three different ways:(a) step motion,(b) square wave and (c) sine wave. As shown in figure \ref{figure6}a (red curve) the step motion definition is equivalent to the use of $A$ to capture the covered length. As we see in figure \ref{figure6}b the square wave length $Sq$ captures $A$ in the vertical direction but also the streamwise length ($U_{\infty}/f$) traversed by TE in one period. Finally the $ \emph{trajectory}$ length  $Tr$ of the sine wave ( see figure \ref{figure6}c) captures the exact $\mathcal{L}$ traversed by the TE over the entire period:

\begin{equation}
T_r= U_{\infty} \int_{0}^{1/f} \sqrt{1 + {[\frac{dy_t (\mathcal{C},t)}{dt}]}^2 } dt 
\end{equation}

The utility of these three metrics is estimated via the agreement (collapse) of the neutral curves for different types of motions in the $f-\mathcal{L}$ domain. Here the $\mathcal{L}$ is normalised by $D$ and this is done so that $f$ can be expressed by $Sr$ at the same domain. Again figures \ref{figure6}a, \ref{figure6}b and \ref{figure6}c show the neutral lines for different kinematics in $\mathcal{L}_D-1/Sr$ domains for the aforementioned types of $\mathcal{L}_D$ namely: $A_D,~\emph{Sq}_D$ and $\emph{Tr}_D$ respectively. The frequency is represented in the inverse form so that a power law between amplitude and frequency is depicted by a straight line. Since previous studies suggest that BvK reversal depends solely on $St_A$, we expect a collapse of the neutral line curves of different kinematics when plotted on an $A_D-1/Sr$ chart. However as seen in figure \ref{figure6}a these curves are heavily dependent on kinematics. On the other hand, as we move towards $Tr$ (which accurately represents the trajectory length) the collapse of the various neutral line curves is significantly improved (see figure \ref{figure6}c). 

\begin{figure}
  \centerline{\includegraphics[width =5 in]{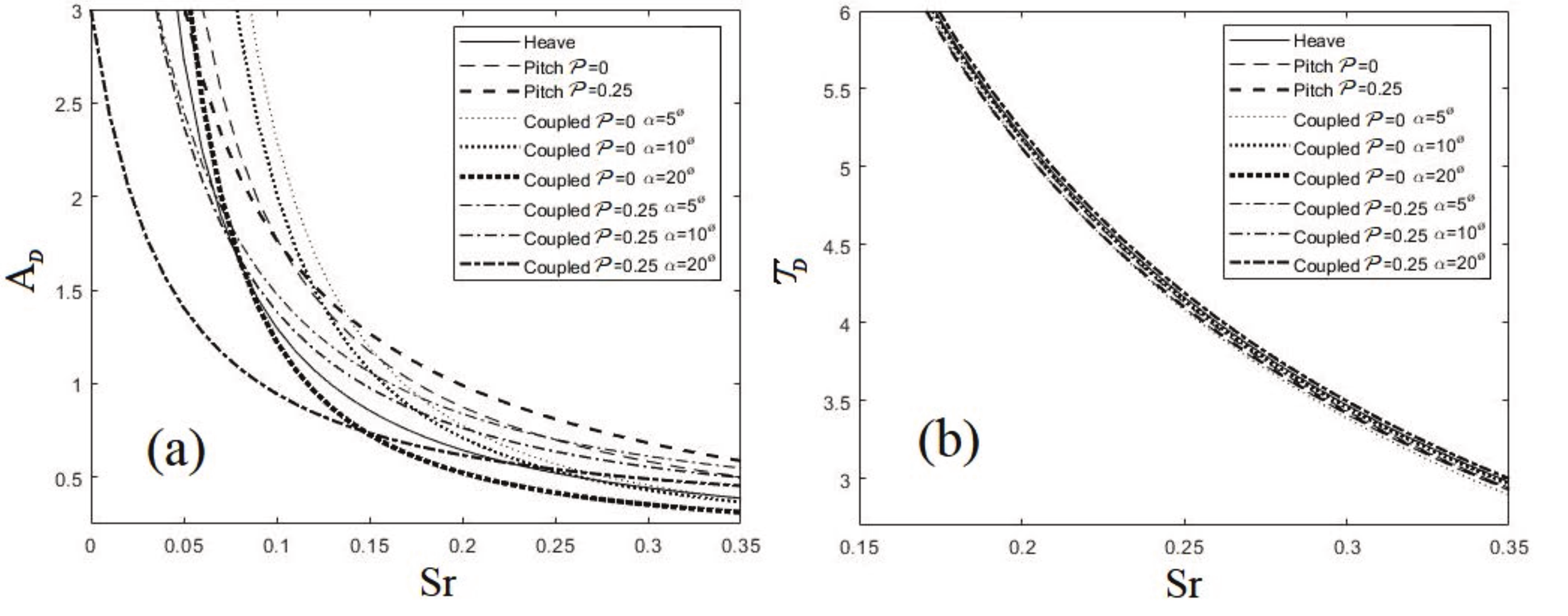}}
\caption{Comparison of neutral line best fit curves for various kinematics on a $Sr-A_D$ phase space (a) and a $Sr-\mathcal{T}_D$ phase space (b) at $Re~=~1173$. Here $\mathcal{T}_D$ is the thickness normalised $\mathcal{T}$. }
\label{figure5}
\end{figure}

$Tr$ is suitable for characterising heave dominant kinematics since the motion of the entire chord can be represented just by the path traversed by the TE. This means that the displaced fluid per cycle of a heave dominant motion can be expressed as $Tr \cdot \mathcal{C}$. However, $Tr$ fails to capture the effects of a chordwise gradient $A$ present in pitch-dominated motions since most of the lateral motion is downstream of the pivot point for $\mathcal{P} <0.5$ (see figure \ref{figure3}b). This means that $Tr$ overestimates the displaced fluid for these particular cases and therefore it is still dependent on kinematics (see figure \ref{figure6}c). The effect of the pitching component can be incorporated into the metric by calculating the $\emph{average}$ trajectory length covered by the entire foil (chord) over one period:
\begin{equation}
\mathcal{T}= \bar{Tr} = U_{\infty} \int_{0}^{1} \int_{0}^{1/f} \sqrt{1+{[\frac{dy_t (s,t)}{dt}]}^2}~dt~ds
\end{equation}  

where, $s$ is the coordinate along the chord with $s = 0$ at LE and $s = 1$ at the TE. 

The $D$-normalised chord averaged trajectory length $\mathcal{T}_D$ is plotted versus $1/Sr$ in figure \ref{figure6}d. The new metric demonstrates remarkable collapse of different kinematics on the curve corresponding to the neutral line of pure heave (where $Tr_D =\mathcal{T}_D$). Interestingly this curve follows the diagonal of a square and as we move towards more inviscid flows it can be expressed as $\mathcal{T}_D \cdot Sr \sim const = 1$. More specifically for $Re=1173$ , $\mathcal{T}_D \cdot Sr \sim 1.035$ while for $Re=1173$ , $\mathcal{T}_D \cdot Sr \sim 1.015$. This product represents the average speed of the foil over one period. Thus the area in the lower right half of the straight line represents the zone where this average speed is less than $U_{\infty}$ while the the upper left is where the speed is larger than $U_{\infty}$. This provides a very simple physical interpretation of the $1/f-\mathcal{T}$ phase space where, in order for the foil to produce thrust, the kinematics has to be tuned in such a way that the chord averaged speed of the foil along the path over one period must be faster than the free stream. In other words, a path length based Strouhal number ($St_{\mathcal{T}} = f \mathcal{T} /U_{\infty}$) should be greater than 1. As the value of $St_{\mathcal{T}}$ increases further beyond 1, the wake of the foil becomes stronger producing higher and higher values of thrust.

\begin{figure}
  \centerline{\includegraphics[width =5 in]{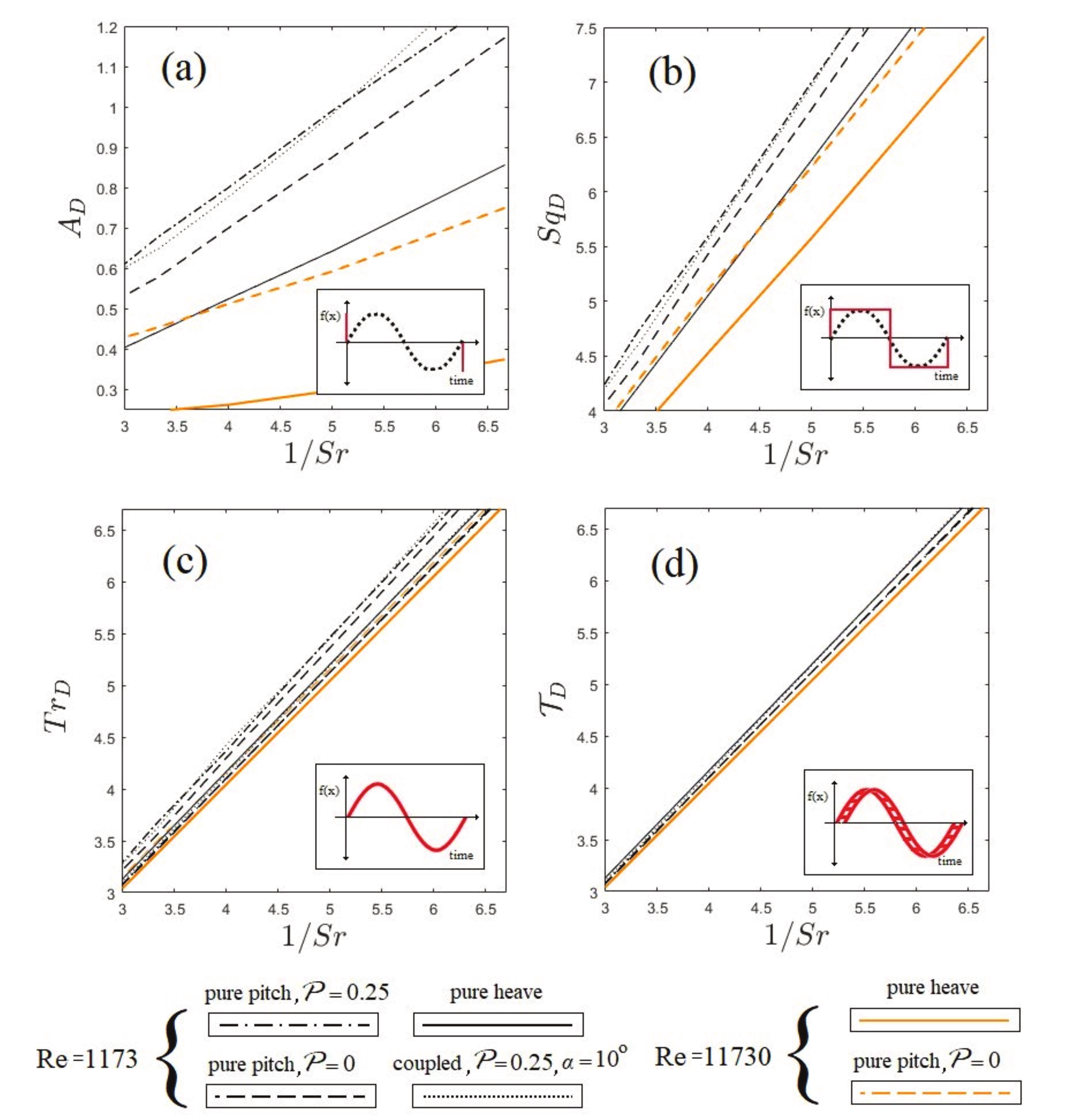}}
\caption{Best fit curves of the neutral line of various harmonic kinematics plotted on $A_D -1/Sr$ (a) , $Sq_D -1/Sr$ (b) , $Tr_D -1/Sr$ (c) and $\mathcal{T}_D -1/Sr$ (d) at $Re~=~1173$ and $Re~=~11730$. Each point on the y-axis expresses the thickness normalised length of the curve (red) of the corresponding approximation of the path traversed by the foil within one cycle.}
\label{figure6} 
\end{figure}
 
The universality of the new length scale is evaluated by further examining the agreement of neutral line curves for various kinematic factors such as $f,~\mathcal{P},~\alpha$. Figure \ref{figure5}b shows all the kinematic options where a collapse was demonstrated. The agreement among these curves deteriorates for $Sr< 0.12$ or $Sr>0.35$ perhaps due to the limitations of using two dimensional numerical simulations. This is consistent with the observations of \cite{Mittal1995} who found that 2D simulations might result in large force fluctuations and erroneous wake patterns. Additionally non periodic wakes have been observed for pure pitch motion at $\mathcal{P}>0.6$ which could also contribute to the poor collapse. As $\alpha$ is essentially an indicator of the heave to the pitch ratio within a coupled motion case it has no real effect on periodicity at least up to $\alpha=20^{\circ}$ and thus the coupled motions presented here agree well with the pure pitch and heave test cases. Clearly the new metric sets a threshold for the drag-to-thrust wake transition of 2D flapping foils for the entire range of kinematics provided that the resultant wake is periodic.

\section{Conclusions}
Two dimensional simulations were conducted for a rigid flapping NACA0016 profile at low and high Reynolds numbers incompressible flow. The wake development towards a reversed BvK street was examined for a variety of harmonic motions, amplitudes and frequencies.
At very low amplitudes and frequencies 2P wake patterns were observed for the coupled motions in agreement with either the pure heave or the pure pitch cases. This is due to the fact that at such low $Sr-A_D$ combinations the coupled motions are dominated by either the heaving or the pitching component. 

In dimensionless Amplitude-Period maps various length scales were evaluated with respect to the neutral line of different motion types. It was revealed that the relationship between $A_D$ and the period is non-linear since the maximum distance from equilibrium cannot properly characterise the displaced volume (or area) required to overcome the drag forces. This is solved by
calculating the length of the path traversed by the foil over one period of oscillation.

On a $1/Sr-Tr_D$ graph the neutral line of pure heave forms a linear curve $y(x)$ where $dy/dx=45^{\circ}$. Since $U_{foil}=Tr_D/(1/Sr)$  this means that thrust is achieved when $U_{foil} > U_{\infty}$. Furthermore the dimensionless chord average trajectories per cycle $\mathcal{T}_D$ of all motion types tested, collapse upon the pure heaving case. In other words the neutral lines of all test cases collapse on a trajectory-based Strouhal $St_{\mathcal{T}} \rightarrow 1$. Thus the new metric can serve as a universal length scale that captures the BvK reversal for every combination of harmonic two dimensional kinematics.

This novel method allows us to to parametrise drag-to-thrust wake transition of a simple two dimensional harmonically oscillating body via the chosen kinematics without the use of complex fluid dynamic equations. Knowing the onset of thrust for a flapping foil via a single parameter can significantly reduce the effort of designing sufficient biomimetic
propulsors. Moreover it will enable scientists and engineers to describe and/or confirm observations regarding the thrust generating strategies and evolution of natural flyers and swimmers. 

\section*{Acknowledgements}
This research was supported financially by the Office of Naval Research award N62909-18-1-2091 and the Engineering and Physical Sciences Research Council doctoral training award [1789955].

\bibliographystyle{jfm}
\bibliography{jfm-instructions}

\begin{thebibliography}{26}
\expandafter\ifx\csname natexlab\endcsname\relax\def\natexlab#1{#1}\fi
\def\au#1{#1} \def\ed#1{#1} \def\yr#1{#1}\def\at#1{#1}\def\jt#1{\textit{#1}}
  \def\bt#1{#1}\def\bvol#1{\textbf{#1}} \def\vol#1{#1} \def\pg#1{#1}
  \def\publ#1{#1}\def\arxiv#1{#1}\def\org#1{#1}\def\st#1{\textit{#1}}

\bibitem[Andersen {\em et~al.\/}(2017)Andersen, Bohr, Schnipper \&
  Walther]{Andersen2017}
{\sc \au{Andersen, A.}, \au{Bohr, T.}, \au{Schnipper, T.} \& \au{Walther,
  J.H.}} \yr{2017}  \at{Wake structure and thrust generation of a flapping foil
  in two-dimensional flow}.  \jt{Journal of Fluid Mechanics}  \bvol{812}.

\bibitem[Anderson {\em et~al.\/}(1998)Anderson, Streitlien, Barrett \&
  Triantafyllou]{Anderson1998}
{\sc \au{Anderson, J.M.}, \au{Streitlien, K.}, \au{Barrett, D.S.} \&
  \au{Triantafyllou, M.S.}} \yr{1998}  \at{Oscillating foils of high propulsive
  efficiency}.  \jt{Journal of Fluid Mechanics}  \bvol{360},  \pg{41--72}.

\bibitem[Birnbaum(1924)]{Birnbaum1924}
{\sc \au{Birnbaum, W.}} \yr{1924}  \at{Das ebene problem des schlagenden
  fl{\"u}gels}.  \jt{ZAMM-Journal of Applied Mathematics and
  Mechanics/Zeitschrift f{\"u}r Angewandte Mathematik und Mechanik}
  \bvol{4}~(4),  \pg{277--292}.

\bibitem[Bohl \& Koochesfahani(2009)]{Bohl2009}
{\sc \au{Bohl, D.G.} \& \au{Koochesfahani, M.M.}} \yr{2009}
  \at{\uppercase{MTV} measurements of the vortical field in the wake of an
  airfoil oscillating at high reduced frequency}.  \jt{Journal of Fluid
  Mechanics}  \bvol{620},  \pg{63--88}.

\bibitem[Cleaver {\em et~al.\/}(2012)Cleaver, Wang \& Gursul]{Cleaver2012}
{\sc \au{Cleaver, D.J.}, \au{Wang, Z.} \& \au{Gursul, I.}} \yr{2012}
  \at{Bifurcating flows of plunging aerofoils at high strouhal numbers}.
  \jt{Journal of Fluid Mechanics}  \bvol{708},  \pg{349--376}.

\bibitem[Fish \& Lauder(2006)]{Fish2006}
{\sc \au{Fish, F.E.} \& \au{Lauder, G.V.}} \yr{2006}  \at{Passive and active
  flow control by swimming fishes and mammals}.  \jt{Annu. Rev. Fluid Mech.}
  \bvol{38},  \pg{193--224}.

\bibitem[Godoy-Diana {\em et~al.\/}(2008)Godoy-Diana, Aider \&
  Wesfreid]{Diana2008b}
{\sc \au{Godoy-Diana, R.}, \au{Aider, J.L.} \& \au{Wesfreid, J.E.}} \yr{2008}
  \at{Transitions in the wake of a flapping foil}.  \jt{Physical Review E}
  \bvol{77}~(1),  \pg{016308}.

\bibitem[Godoy-Diana {\em et~al.\/}(2009)Godoy-Diana, Marais, Aider \&
  Wesfreid]{Diana2008a}
{\sc \au{Godoy-Diana, R.}, \au{Marais, C.}, \au{Aider, J.L.} \& \au{Wesfreid,
  J.E.}} \yr{2009}  \at{A model for the symmetry breaking of the reverse
  b{\'e}nard--von k{\'a}rm{\'a}n vortex street produced by a flapping foil}.
  \jt{Journal of Fluid Mechanics}  \bvol{622},  \pg{23--32}.

\bibitem[Koochesfahani(1989)]{Koochesfahani1989}
{\sc \au{Koochesfahani, M.M.}} \yr{1989}  \at{Vortical patterns in the wake of
  an oscillating airfoil}.  \jt{AIAA journal}  \bvol{27}~(9),  \pg{1200--1205}.

\bibitem[Maertens \& Weymouth(2015)]{Maertens2015}
{\sc \au{Maertens, A.P.} \& \au{Weymouth, G.D.}} \yr{2015}  \at{Accurate
  cartesian-grid simulations of near-body flows at intermediate reynolds
  numbers}.  \jt{Computer Methods in Applied Mechanics and Engineering}
  \bvol{283},  \pg{106 -- 129}.

\bibitem[Mittal \& Balachandar(1995)]{Mittal1995}
{\sc \au{Mittal, R.} \& \au{Balachandar, S.}} \yr{1995}  \at{Effect of
  three‐dimensionality on the lift and drag of nominally two‐dimensional
  cylinders}.  \jt{Physics of Fluids}  \bvol{7}~(8),  \pg{1841--1865}.

\bibitem[Platzer \& Jones(2008)]{Platzer2008}
{\sc \au{Platzer, M.} \& \au{Jones, K.}} \yr{2008} Flapping wing
  aerodynamics-progress and challenges.  \bt{In {\em 44th AIAA Aerospace
  Sciences Meeting and Exhibit\/}},  \pg{p. 500}.

\bibitem[Polet {\em et~al.\/}(2015)Polet, Rival \& Weymouth]{Polet2015}
{\sc \au{Polet, D.T.}, \au{Rival, D.E.} \& \au{Weymouth, G.D.}} \yr{2015}
  \at{Unsteady dynamics of rapid perching manoeuvres.}  \jt{Journal of Fluid
  Mechanics}  \bvol{767},  \pg{323–341}.

\bibitem[Ramamurti \& Sandberg(2001)]{Ramamurti2001}
{\sc \au{Ramamurti, R.} \& \au{Sandberg, W.}} \yr{2001} Computational study of
  \uppercase{3D} flapping foil flows.  \bt{In {\em 39th Aerospace Sciences
  Meeting and Exhibit\/}},  \pg{p. 605}.

\bibitem[Read {\em et~al.\/}(2003)Read, Hover \& Triantafyllou]{Read2003}
{\sc \au{Read, D.A.}, \au{Hover, F.S.} \& \au{Triantafyllou, M.S.}} \yr{2003}
  \at{Forces on oscillating foils for propulsion and manoeuvering}.
  \jt{Journal of Fluids and Structures}  \bvol{17}~(1),  \pg{163--183}.

\bibitem[Streitlien \& Triantafyllou(1998)]{Streitlien1998}
{\sc \au{Streitlien, K.} \& \au{Triantafyllou, G.S.}} \yr{1998}  \at{On thrust
  estimates for flapping foils}.  \jt{Journal of fluids and structures}
  \bvol{12}~(1),  \pg{47--55}.

\bibitem[Taylor {\em et~al.\/}(2003)Taylor, Nudds \& Thomas]{Taylor2003}
{\sc \au{Taylor, G.K.}, \au{Nudds, R.L.} \& \au{Thomas, A.L.R.}} \yr{2003}
  \at{Flying and swimming animals cruise at a \uppercase{S}trouhal number tuned
  for high power efficiency}.  \jt{Nature}  \bvol{425}~(6959),  \pg{707}.

\bibitem[Thiria {\em et~al.\/}(2006)Thiria, Goujon-Durand \&
  Wesfreid]{Thiria2006}
{\sc \au{Thiria, B.}, \au{Goujon-Durand, S.} \& \au{Wesfreid, J.E.}} \yr{2006}
  \at{The wake of a cylinder performing rotary oscillations}.  \jt{Journal of
  Fluid Mechanics}  \bvol{560},  \pg{123--147}.

\bibitem[Triantafyllou {\em et~al.\/}(1993)Triantafyllou, Triantafyllou \&
  Grosenbaugh]{Triantafyllou1993}
{\sc \au{Triantafyllou, G.S.}, \au{Triantafyllou, M.S.} \& \au{Grosenbaugh,
  M.A.}} \yr{1993}  \at{Optimal thrust development in oscillating foils with
  application to fish propulsion}.  \jt{Journal of Fluids and Structures}
  \bvol{7}~(2),  \pg{205--224}.

\bibitem[Triantafyllou {\em et~al.\/}(2004)Triantafyllou, Techet \&
  Hover]{Triantafyllou2004}
{\sc \au{Triantafyllou, M.S.}, \au{Techet, A.H.} \& \au{Hover, F.S.}} \yr{2004}
   \at{Review of experimental work in biomimetic foils}.  \jt{IEEE Journal of
  Oceanic Engineering}  \bvol{29}~(3),  \pg{585--594}.

\bibitem[Triantafyllou {\em et~al.\/}(1991)Triantafyllou, Triantafyllou \&
  Gopalkrishnan]{Triantafyllou1991}
{\sc \au{Triantafyllou, M.S.}, \au{Triantafyllou, G.S.} \& \au{Gopalkrishnan,
  R.}} \yr{1991}  \at{Wake mechanics for thrust generation in oscillating
  foils.}  \jt{Physics of Fluids A: Fluid Dynamics}  \bvol{3}~(12),
  \pg{2835–--2837}.

\bibitem[Vial {\em et~al.\/}(2004)Vial, Bellon \& Hern{\'a}ndez]{Vial2004}
{\sc \au{Vial, M.}, \au{Bellon, L.} \& \au{Hern{\'a}ndez, R.H.}} \yr{2004}
  \at{Mechanical forcing of the wake of a flat plate}.  \jt{Experiments in
  Fluids}  \bvol{37}~(2),  \pg{168--176}.

\bibitem[Von~Karman(1935)]{vonKarman1935}
{\sc \au{Von~Karman, T.}} \yr{1935}  \at{General aerodynamic theory-perfect
  fluids}.  \jt{Aerodynamic theory}  \bvol{2},  \pg{346--349}.

\bibitem[Wang(2005)]{Wang2005}
{\sc \au{Wang, Z.J.}} \yr{2005}  \at{Dissecting insect flight}.  \jt{Annual
  Review of Fluid Mechanics}  \bvol{37}~(1),  \pg{183--210},  \arxiv{arXiv:
  https://doi.org/10.1146/annurev.fluid.36.050802.121940}.

\bibitem[Weymouth \& Yue(2011)]{Weymouth2011}
{\sc \au{Weymouth, G.D.} \& \au{Yue, D.K.P.}} \yr{2011}  \at{Boundary data
  immersion method for cartesian-grid simulations of fluid-body interaction
  problems}.  \jt{Journal of Computational Physics}  \bvol{230}~(16),  \pg{6233
  -- 6247}.

\bibitem[Williamson \& Roshko(1988)]{Williamson1988}
{\sc \au{Williamson, C.H.K.} \& \au{Roshko, A.}} \yr{1988}  \at{Vortex
  formation in the wake of an oscillating cylinder}.  \jt{Journal of Fluids and
  Structures}  \bvol{2}~(4),  \pg{355 -- 381}.

\end{thebibliography}

\end{document}